Taylor & Francis by informa

REVIEW ARTICLE

OPEN ACCESS Check for updates

# Clinical predictions using artificial intelligence in neurosurgery and its subspecialties: a systematic review and meta-analysis

Prajwal Ghimire[a,b,c], Aswin Chari[d,e], Jonathan Shapey[a,b], and Keyoumars Ashkan[a,b,f]

[a]Department of Neurosurgery, King's College Hospital NHS Foundation Trust, London, UK; [b]School of Biomedical Engineering & Imaging Sciences, King's College London, London, UK; [c]King's Institute of Artificial Intelligence, King's College London, London, UK; [d]Department of Neurosurgery, Great Ormond Street Hospital for Children NHS Foundation Trust, London, UK; [e]Institute of Child Health, University College London, London, UK; [f]Institute of Psychiatry, Psychology & Neuroscience, King's College London, London, UK

**ABSTRACT**

**Introduction:** Artificial intelligence (AI) models for clinical prediction are proliferating across neurosurgery, but their aggregate discriminative performance, distribution along the care pathway and methodological quality are unclear. We systematically reviewed and meta-analysed these studies in neurosurgery and its subspecialities along with their risk of bias appraisal.

**Methods:** We performed a PRISMA search on PubMed and Ovid for peer-reviewed original studies (2015–2026) using AI to predict a clinical outcome in neurosurgical populations. Metrics (primary- AUC; secondary- accuracy, sensitivity, specificity, c-index) were pooled using random effects modelling, overall, by subspeciality and care-pathway stage. Each study was classified as diagnostic, prognostic or forecasting and was assigned an evidence level. Clinical prediction task sub-stratification analysis was performed. Risk-of-bias was assessed using PROBAST. PROSPERO registered: CRD420261436614

**Results:** 1063 studies were included. Across 451 (42.2%) studies reporting AUC, pooled discrimination was 0.86 (95% CI: 0.849–0.862), ranging from 0.84 (spine) to 0.87 (neuro-oncology), with extreme heterogeneity ($I^2$ 98.6%, $\tau^2$=0.27; 95% prediction interval: 0.68–0.94). Performance varied by clinical pathway stage, highest for diagnosis, grading and molecular tasks and lowest for post-operative tasks; pre-diagnosis, pre-operative and intra-operative applications were sparse, and treatment models were confined to neuro-oncology. Only 51 studies (4.8%) were using prospective data with external validation, whereas 619 (58%) rested on internal validation alone or no validation. 89% studies were at high risk-of-bias, driven by the analysis domain. Pooled discrimination was unchanged (AUC: 0.86%) when restricted to externally validated or non-high-risk studies.

**Conclusion:** AI models report high performance across all the neurosurgical subspecialities, concordant across AUC, accuracy, sensitivity and specificity and robust to risk-of-bias stratification. Performance is highest for imaging-anchored diagnostic tasks and lowest at decision-facing post-operative stage. However, the literature is dominated by optimism-prone, internally-validated and poorly calibrated models with extreme heterogeneity. External validation, calibration analysis and reporting guidelines adherence remain prerequisites for clinical adoption.

## 1. Introduction

Clinical prediction is key in neurosurgical decision-making: estimating the probability of a molecular tumour subtype, of haemorrhage or rupture, of seizure freedom after epilepsy surgery, or of complications and functional recovery after spinal surgery. Over the past decade, artificial intelligence (AI) methods have been applied to these tasks at rapidly increasing scale, ranging from logistic regression, tree ensembles, support-vector machines, radiomics pipelines and deep neural networks.[1–4]

Individual studies frequently report excellent discrimination, but enthusiasm has outpaced methodological scrutiny. The wider clinical machine learning

CONTACT Prajwal Ghimire prajwal.1.ghimire@kcl.ac.uk | prajwal.ghimire@nhs.net School of Biomedical Engineering & Imaging Sciences, King's College London, London, UK.

(ML) literature is characterised by small samples, internal-only validation, absent calibration and incomplete reporting- features that inflate apparent performance and limit generalisability of these prediction models.[5,6] Whether neurosurgical AI prediction models share these weakness and how performance and bias vary across subspecialities, has not been comprehensively quantified.

We therefore performed a systematic review and meta-analysis of AI prediction studies across the neurosurgical subspecialities. The aims were (i) to map the machine learning methods and clinical prediction tasks in use, (ii) to pool discriminative performance, (iii) to map the application of these tools in the clinical care pathway- from pre-diagnosis to adjuvant/neo-adjuvant treatment; (iv) to attempt classification of models, define clinical translation stage and explore level of evidence (v) to appraise risk of bias and applicability.

## 2. Methods

The review followed the Preferred Reporting Items for Systematic Reviews and Meta-analyses[7,8] and incorporated the developing PRISMA-AI[9] reporting considerations for studies of machine learning models. A PROSPERO registration was performed (CRD420261436614).

### 2.1. Eligibility criteria

We included original peer-reviewed studies that applied AI or ML as part of the methodology to predict a clinical outcome in neurosurgical populations across its subspecialities: neuro-oncology, neuro-vascular, spine, functional/epilepsy, skull-base, paediatrics, general and trauma. Retrospective, prospective and trial designs were eligible.

### 2.2. Exclusion criteria

We excluded reviews, editorials, letters, conference proceedings, non-human studies and studies without a clinically relevant predictive outcome. Title-level review articles were removed during screening.

### 2.3. Search and selection

PubMed and a multi-field OVID (Embase, Medline) databases were searched for the period of 2015–2026 using medical subject headings (MeSH) terms. A low precision high sensitivity search was conducted using subject headings and exploding terms. Records were deduplicated and screened against a neurosurgical relevance and study-type to remove reviews and non-neurosurgical records. All results were reviewed with dual-reviewer adjudication.

### 2.4. Statistical analysis

Study-level discrimination (area-under-the-curve, AUC) was logit-transformed and pooled using random-effects modelling,[10] overall and within each subspeciality. The standard error of each logit-AUC was derived from the reported 95% confidence intervals (CI); where unavailable, it was approximated from sample size using the Hanley-McNeil method.[11] Statistical heterogeneity[10] was summarised with meta-analysis metrics $I^2$ and $\tau^2$, and 95% prediction intervals were calculated; estimates were back-transformed to the AUC scale. Pre-specified sensitivity analyses restricted pooling to externally validated studies and to studies not rated high-risk of bias.

Secondary performance metrics (accuracy, sensitivity, specificity, c-index) were pooled with the same random-effects model on the logit scale. Standard errors were derived from the reported sample size using the binomial approximation or imputed as the median where sample size was not available. Secondary metrics were additionally pooled by subspeciality and by clinical pathway stage, reporting only strata containing at least five studies.

To explore heterogeneity, subgroup analyses based on subspeciality, clinical pathway stage, primary model family, validation status, level of evidence, overall risk of bias, predicted outcome, study design and source of standard error was performed with pooling within each level and testing level means differences using Cochran's Q between groups. Random-effects meta-regression was fitted with methods-of-moments estimation of the $\tau^2$ metric, univariably and in a multivariate analysis accounting subspeciality, validation status, model family, sample size and publication year.

The distribution of performance metric was summarised by threshold band (<0.70, 0.70–0.80, 0.80–0.90, ≥0.90) overall, by subspeciality and by clinical pathway stage. Small-study effects were examined with funnel plots and with Egger's regression test for funnel asymmetry.

Proportion of studies reporting calibration were tabulated.

### 2.5. Risk-of-bias assessment

Risk of bias and applicability were assessed with PROBAST[12] across its four domains (participants,

predictors, outcome, analysis) and 21 signalling questions, incorporating PROBAST-AI[13] considerations for ML models. Reporting was assessed against TRIPOD[14] and TRIPOD-AI extension.[15]

### 2.6. Data extraction

Extraction was structured against the CHARMS checklist[16] for systematic reviews of prediction modelling studies covering its domain of source of data, participants, outcome to be predicted, candidate predictors, sample size, missing data, model development, model performance, model evaluation, results and interpretation. Following CHARMS strategy, each study was classified as a diagnostic (estimating the probability that a condition is currently present including grade, subtype, molecular biomarkers from imaging), prognostic (estimating probability of a future event) or a forecasting model (dynamic or time-updated prediction over a temporal set of events).

### 2.7. Stage of clinical translation

Each study was placed on a six-level translation ladder based on CHARMS, TRIPOD-AI[15] and DECIDE-AI[17] guidelines: T0 development only; T1 externally or temporally validated; T2 a usable artefact released (web-based, application or published nomogram); T3 prospective clinical evaluation of the model; T4 an impact or implementation study; T5 regulatory clearance or documented routine deployment.

### 2.8. Level of evidence

No established evidence hierarchy grades clinical prediction model studies by validation status. We therefore descriptively summarised the studies, informed by the Oxford Centre for Evidence-Based Medicine (OECBM)[18] design with combination of study design and validation status: prospective cohort or trial data with external or temporal validation; prospective or trial design without external validation or retrospective or unspecified design with external validation; internal validation only (resampling, cross-validation or split-sample); no validation reported (apparent performance only). Studies were classified into by TRIPOD[14,15] model-evidence type (Type 1a, development only; Type 1b, development with resampling; Type 2b, non-random split-sample; Type 3, development with validation on separate data; Type 4, validation only of an existing model). Validation status was taken from the PROBAST[12] signalling answers derived from each article.

### 2.9. Clinical pathway-stage classification

Each study was assigned to a single stage according to its title, model and outcome in order to describe the stage in clinical pathway each model operates in: pre-diagnosis/screening/risk, diagnosis/grading/molecular, pre-operative, operative/intra-operative, post-operative, baseline prognostication, adjuvant/neo-adjuvant treatment and unclassified/multi-stage residual.

### 2.10. Clinical prediction task sub-stratification

Studies were identified within specific clinical predictions tasks where the neurosurgical population, the neurosurgical predictors and the predicted outcome were comparable. Discrimination was pooled with random effects modelling on the logit-AUC scale where at least 5 studies reported an AUC for each subspeciality. Other studies (<5) were not pooled, but AUC, median and range were reported.

## 3. Results

1063 studies met eligibility criteria, spanning 2015–2026 (Figure 1). The studies were grouped as per their subspeciality: neuro-oncology ($n = 470$, 44%); spine ($n = 184$); neuro-vascular ($n = 180$); functional/epilepsy ($n = 83$); trauma ($n = 82$), paediatrics ($n = 26$), skull-base ($n = 24$) and general neurosurgery ($n = 14$) (Table 1). Reported designs were predominantly retrospective. Specific algorithms in the model architecture were identified: logistic regression ($n = 421$, 40%), deep learning/neural networks ($n = 289$, 27%), LASSO-regularised regression($n = 268$, 25%), random forest ($n = 256$, 24%); support-vector machines ($n = 244$, 23%), Cox regression ($n = 228$, 21%) and gradient boosting ($n = 174$, 16%); transformer ($n = 18$, 2%), foundation-model ($n = 8$, 0.8%) approaches remained rare. Studies commonly compared several algorithms ($n = 521$, 49% $\geq$ 2 models) (Figure 2).

### 3.1. Pooled discrimination

Across 451 studies reporting AUC, pooled discrimination was high and similar across all subspecialities, spanning 0.84–0.92: 0.87 (0.86–0.88, $n = 172$) in neuro-oncology; 0.84 (0.82–0.86, $n = 83$) in neuro-vascular; 0.84 (0.83–0.85, $n = 94$) in spine; 0.87 (0.83–0.91, $n = 11$) in skull-base; 0.84 (0.73–0.91, $n = 27$) in

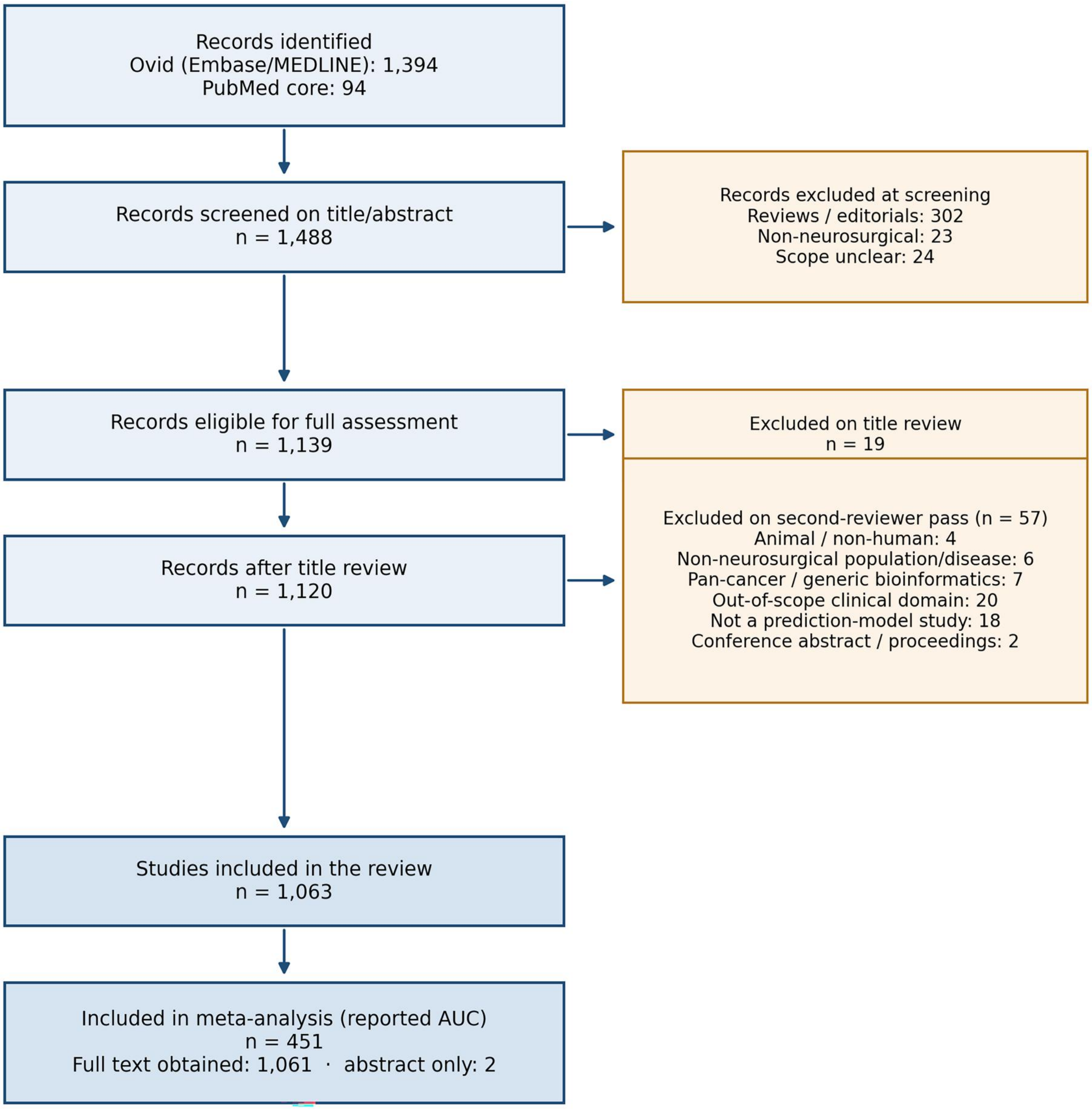

**Figure 1.** PRISMA-based search strategy of the systematic review for AI-based clinical prediction tools in neurosurgery. AUC: area under the curve

**Table 1.** Included studies by subspeciality and AUC reporting.

| Subspeciality | Studies, n(%) | Reporting AUC, n |
|---|---|---|
| Neuro-oncology | 470 (44.2%) | 172 |
| Spine | 184 (17.3%) | 94 |
| Neuro-vascular | 180 (16.9%) | 83 |
| Functional/Epilepsy | 83 (7.8%) | 27 |
| Trauma | 82 (7.7%) | 46 |
| Pediatric neurosurgery | 26 (2.5%) | 10 |
| Skull-base | 24 (2.3%) | 11 |
| General | 14 (1.3%) | 8 |
| Total | 1063 (100%) | 451 |

AUC: Area under the curve.

functional/epilepsy; 0.85 (0.79–0.90, $n = 10$) in paediatrics; 0.86 (0.83–0.88, $n = 46$) in trauma; 0.92 (0.88–0.94, $n = 8$) in general neurosurgery; overall pooled AUC 0.86 (0.85–0.86) (Table 2 and Figure 3). Heterogeneity was extreme in every analysis ($I^2$ 76.1-99.1% by subspeciality; 98.6% overall) and the corresponding between-study variance was $\tau^2$=0.27 implying that the overall 95% prediction interval was wide (0.68–0.94) and that the performance in a new setting is less certain than the CI and $I^2$ implies. Per-study forest plots by subspeciality and by clinical pathway stage are plotted in Suppl. Figures S1 and S2. 7.8% of these studies ($n = 35$) had AUC <0.70 and 39.5% studies ($n = 178$) reported AUC range of 0.80–0.90 with neuro-oncology ($n = 67$), neurovascular ($n = 35$), spine ($n = 42$) and skull base ($n = 4$) subspeciality with 36.4% to 44.7% studies in the AUC range of 0.80–0.90. Distribution of the AUCs by threshold band and subspeciality are demonstrated in Suppl. Table S1.

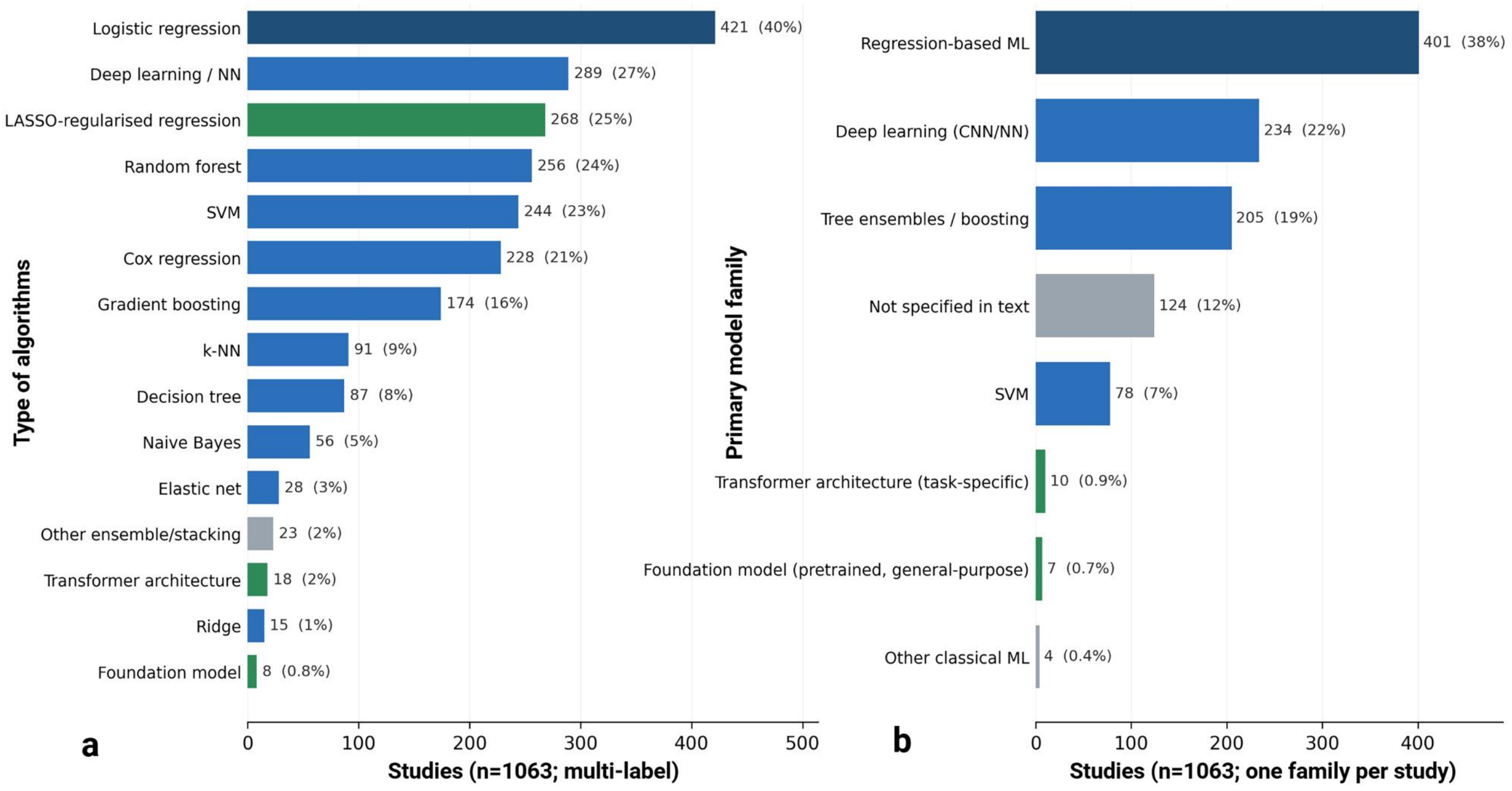

**Figure 2.** Machine learning algorithms identified in the studies (a) and primary model family (b) across the studies.
K-NN: K nearest neighbour, SVM: support vector machine, ML: machine learning; CNN: convolutional neural network; NN: neural network; LASSO: Least absolute shrinkage and selection operator

**Table 2.** Random-effects pooled AUC, heterogeneity and prediction intervals.

| Subspeciality | n | Pooled AUC (95% CI) | $I^2$ (%) | $\tau^2$ | 95% PI |
|---|---|---|---|---|---|
| Neuro-oncology | 172 | 0.87 (0.86–0.88) | 95.6 | 0.595 | 0.59-0.97 |
| Spine | 94 | 0.84 (0.83–0.85) | 99.1 | 0.118 | 0.72-0.91 |
| Neuro-vascular | 83 | 0.84 (0.82–0.86) | 97.7 | 0.415 | 0.60-0.95 |
| Trauma | 46 | 0.86 (0.83–0.88) | 97.7 | 0.450 | 0.60-0.96 |
| Functional/epilepsy | 27 | 0.84 (0.73–0.91) | 98.8 | 2.681 | 0.14-0.99 |
| Skull Base | 11 | 0.87 (0.83–0.91) | 76.1 | 0.262 | 0.67-0.96 |
| Paediatrics | 10 | 0.85 (0.79–0.90) | 85.7 | 0.364 | 0.57-0.96 |
| General | 8 | 0.92 (0.88–0.94) | 90.2 | 0.333 | 0.71-0.98 |
| Overall | 451 | 0.86 (0.85–0.86) | 98.6 | 0.264 | 0.68–0.94 |

AUC: area under the curve; CI: confidence interval; PI: prediction intervals; *n*: number of studies; $I^2$, $\tau^2$: statistical heterogeneity metrics.

## 3.2. Prediction across the clinical pathway

All the studies were mapped onto the clinical neurosurgical care pathway (Figure 4). The majority of studies were dedicated to diagnosis and prognostication: 276 studies addressed diagnosis, grading or molecular characterisation and 361 studies for baseline prognostication (total: 60% of studies), followed by the post-operative stage ($n = 238$), where spine, trauma and functional/epilepsy work is concentrated. Applications at the pathway extremes were sparse: pre-diagnosis/screening ($n = 32$, predominantly neurovascular aneurysm rupture-risk), pre-operative decision support ($n = 25$) and intra-operative use ($n = 22$) and adjuvant/neoadjuvant treatment-response modelling ($n = 51$) was almost entirely confined to neuro-oncology ($n = 47$ out of 51). Median reported discrimination was highest at the imaging-anchored diagnosis/molecular stage (0.90) and lowest at post-operative (0.82) (Suppl. Figure S3, Suppl. Table S2).

## 3.3. Secondary performance metrics

Secondary metrics (Table 3) pooled with the same random-effects model were: accuracy 0.88 (95% CI 0.86–0.90; $n = 83$), specificity 0.88 (0.84–0.90; $n = 64$), sensitivity 0.87 (0.85–0.89; $n = 57$) and c-index 0.80 (0.77–0.83; $n = 31$)(Suppl. Figure S4a, Suppl. Table S3a). Sub-group analyses ($n \geq 5$ per stratum; Suppl. Figure S4b,S4c, Suppl. Table S3b,S3c) reproduced the pathway gradient seen for AUC: accuracy(0.89), sensitivity(0.89) and specificity(0.93) were all highest at the diagnosis/grading/molecular stage and lower at baseline prognostication and post-operatively.

## 3.4. Types of prediction models

694 studies (65%) developed prognostic models, 283 (27%) diagnostic models and only 8 forecasting models. Diagnostic models pooled higher performance than prognostic models (0.89, 95% CI: 0.88–0.90, versus 0.83, 0.83–0.84) consistent with the observation that distinguishing a clinical state that already exists is an easier task compared to predicting a future clinical event. Supplemental Table S4 describes the prediction model types by subspeciality with pooled metrics.

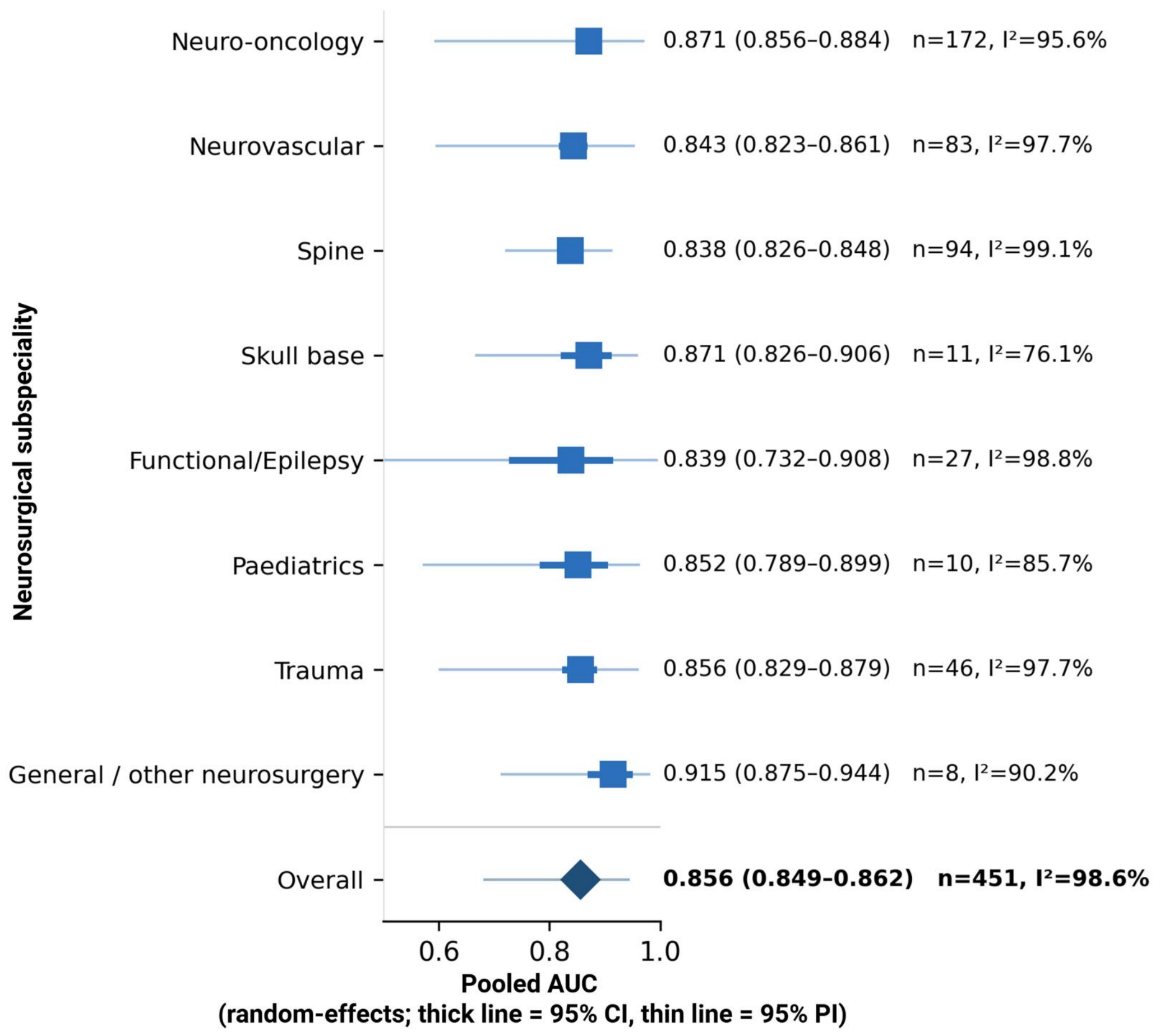

**Figure 3.** Pooled discrimination metric (AUC) by neurosurgical subspeciality.
*AUC: area under the curve; CI: confidence interval, PI: prediction interval; $I^2$: statistical heterogeneity metric*

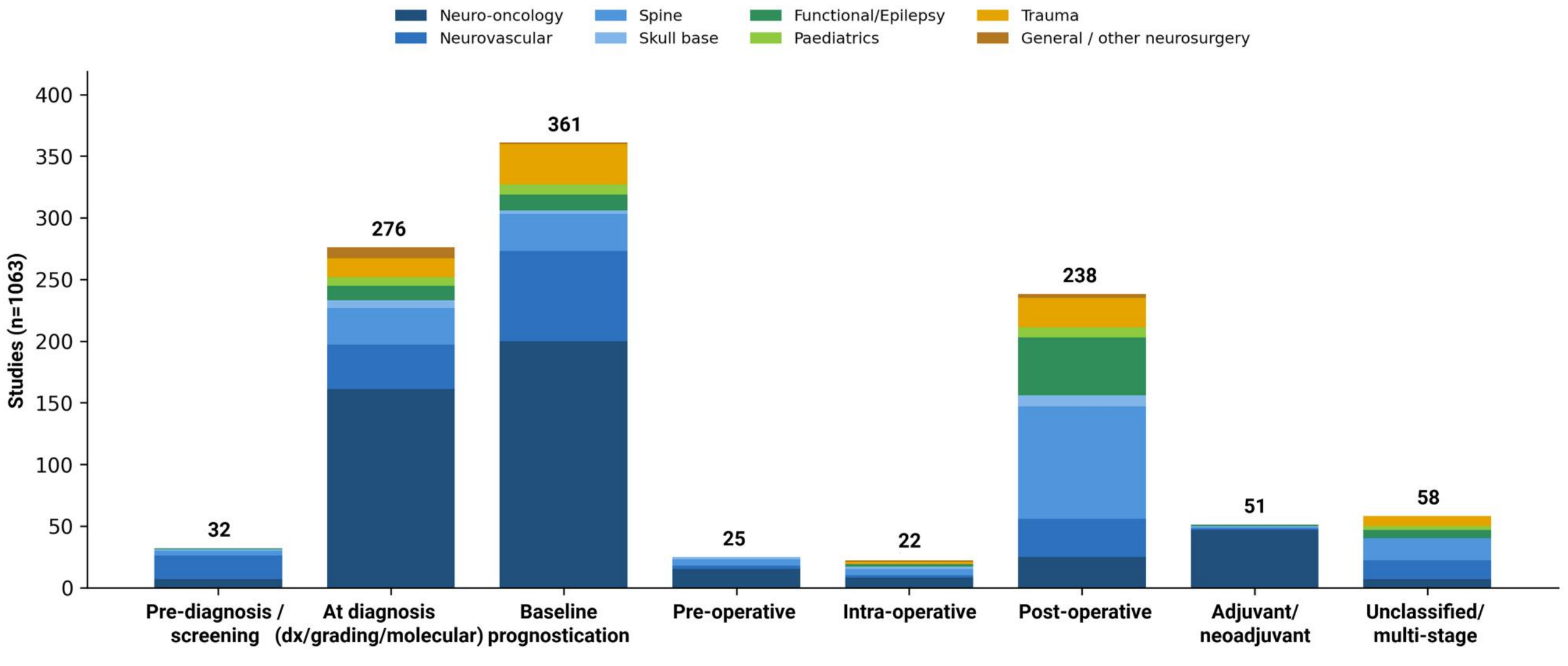

**Figure 4.** Machine learning prediction studies in neurosurgery by clinical-pathway stage and subspeciality.
dx: diagnostic

### *3.5. Level of evidence*

By TRIPOD[14,15] study type, 161 studies (15%) reported apparent performance only (Type 1a), 420 (40%) used resampling (Type 1b), 123 (12%) used a random split-sample design (Type 2a), 20 (2%) used a non-random split: temporal, geographic or with randomisation not stated (Type 2b), 336 (32%) validated

**Table 3.** Pooled secondary performance metrics, overall and by clinical pathway stage.

| Metric | Clinical pathway stage / stratum | n | Pooled estimate (95% CI) | $I^2$ (%) |
|---|---|---|---|---|
| Accuracy | Overall | 83 | 0.88 (0.86–0.90) | 98.6 |
| | At diagnosis: dx/grading/molecular | 41 | 0.89 (0.86–0.92) | 97.5 |
| | Baseline prognostication | 7 | 0.84 (0.73–0.91) | 97.7 |
| | Post-operative | 26 | 0.88 (0.84–0.91) | 99.2 |
| Sensitivity | Overall | 57 | 0.87 (0.85–0.89) | 98.2 |
| | At diagnosis: dx/grading/molecular | 22 | 0.89 (0.83–0.93) | 98.8 |
| | Baseline prognostication | 10 | 0.85 (0.80–0.89) | 92.6 |
| | Post-operative | 16 | 0.87 (0.84–0.90) | 97 |
| Specificity | Overall | 64 | 0.88 (0.84–0.90) | 99.2 |
| | At diagnosis: dx/grading/molecular | 24 | 0.93 (0.88–0.96) | 99 |
| | Baseline prognostication | 16 | 0.81 (0.75–0.86) | 96.2 |
| | Post-operative | 16 | 0.83 (0.75–0.89) | 98.9 |
| c-index | Overall | 31 | 0.80 (0.77–0.83) | 92.7 |
| | Baseline prognostication | 23 | 0.79 (0.76–0.83) | 93.9 |

Strata with $n \geq 5$; CI: confidence interval; $I^2$: statistical heterogeneity metric; *n*: number of studies.

on separate data (Type 3) and only 2 were validation-only studies of an existing model (Type 4)(Table 4). Level of evidence differed by subspeciality: neuro-vascular had the highest proportion of prospective studies with external/temporal validation (10%) and neuro-oncology the lowest (2%). Level of evidence tracked risk of bias closely: 100% of no validation studies and 93% of internal validation only studies were a high overall risk of bias, compared with 82% of prospective studies with external or temporal validation. CHARMs model evaluation of validation methods is described in Suppl. Table S5. Only 51 studies (4.8%) reached prospective or trial data with external or temporal validation. 479 (45%) internal validation only and 140 (13.2%) no validation.

### 3.6. Stage of clinical translation

Only 4.9% of all the studies were at the stage of clinical use. 679 studies (63.9%) only developed models and went no further; 332 (31.2%) reached external or temporal validation; 34 (3.2%) released a useable artefact such as a web-based application; 15 (1.4%) reported prospective clinical evaluation of the model. No studies reported an impact or implementation study and only 3 (0.3%) studies reported regulatory clearance or documented routine deployment (Suppl. Table S6).

### 3.7. Risk of bias

Overall risk of bias was high in 962 of 1063 studies (90%) and unclear in 101; none reached low overall risk of bias. The analysis domain was the dominant driver (high-960 studies), reflecting absent calibration assessment, reliance on internal or cross-validation without external/temporal validation and uncorrected optimism. The participants domain was comparatively reassuring (low-674 studies), whereas the predictor and outcome domains were frequently unclear owing to incomplete reporting (Figure 5a). Sub-speciality risk of bias by domain remained high in each group (range: 88%-96%)( Figure 5b). Reporting completeness by each study derived from PROBAST-AI signalling answers are plotted in Suppl. Figure S5.

### 3.8. Sensitivity analyses

Pooled discrimination was stable across risk-of-bias strata. When restricted to externally or temporally validate studies ($n = 241$), pooled AUC was 0.86 (95% CI 0.85–0.86) and excluding high-risk-of-bias studies ($n = 71$) pooled AUC was 0.86 (0.84–0.89), both essentially identical to the primary estimate of 0.86 (Figure 5c). The high pooled discrimination is therefore not an artefact of the weakest studies; however, the better-conducted subset did not resolve the extreme heterogeneity ($I^2$ remained > 98%).

### 3.9. Heterogeneity-source and calibration analysis

Extreme heterogeneity was observed in each analysis. The sources of heterogeneity was examined in sub-group analyses (Suppl. Table S7) where subgroup mean differed significantly for 6 of the 9 sources tested: subspeciality ($Q = 22.73$, df = 7, $p = 0.002$); clinical pathway stage ($Q = 113.21$, df = 7, $p < 0.001$); primary model family ($Q = 21.24$, df = 5, $p < 0.001$); validation status ($Q = 8.39$, df = 2, $p = 0.015$); predicted outcome ($Q = 108.24$, df = 4, $p < 0.001$); study design ($Q = 15.53$, df = 3, $p = 0.001$). Overall risk of bias ($Q = 65$, $p = 0.42$) and source of standard error ($p = 0.19$) did not discriminate. In random-effects meta-regression the largest single contributor was sub-speciality, explaining 24.9% of $\tau^2$ with validation status explaining 6.2%, sample size 3.6% and risk of bias 1.7%. None of the other sources explained $\tau^2$. This difference likely arose from large within-stratum

**Table 4.** Level of evidence based on study design with validation and TRIPOD prediction-model study type (total 1063 studies).

| Category based on design and validation | Studies, n (%) | Category based on TRIPOD | Definition | Studies, n (%) |
|---|---|---|---|---|
| Prospective cohort or trial data with external/temporal validation | 51 (4.8%) | Type 1a | Development only (apparent performance) | 161 (15.1%) |
| Prospective cohort/trial without external validation, OR retrospective/unspecified design with external validation | 393 (37%) | Type 1b | Development + resampling (internal validation) | 420 (39.5%) |
| Internal validation only (resampling, cross-validation or split-sample) | 479 (45%) | Type 2a | Random split-sample development and validation | 123 (11.6%) |
| No validation reported (apparent performance only) | 140 (13.2%) | Type 2b | Non-random split-sample (temporal/geographic or randomisation not stated) | 20 (1.9%) |
| | | Type 3 | Development + validation on separate data | 336 (31.6%) |
| | | Type 4 | Validation only of an existing model | 2 (0.2%) |
| | | unclear | Validation strategy not ascertainable from text | 1 (0.1%) |

TRIPOD: Transparent Reporting of a multivariable prediction model for Individual Prognosis Or Diagnosis.

heterogeneity with marked variability between strata (range of $\tau^2$: 0.09–1.04 across pathway stages). A multivariate model combining subspeciality, validation status, model family, sample size and year explained 11.8% of $\tau^2$ ($n = 356$), having a residual $\tau^2$ of 0.209 against a total of 0.264.

Calibration, on the other hand, was identifiable only in 36.2% studies overall ($n = 385$) and in 45.5% studies ($n = 205$) of the 451 studies reporting AUC. Decision-curve analysis appeared in 190 (17.9%) studies. Calibration reporting was sparse and heterogenous.

### 3.10. Small-study effects and publication bias analysis

The AUC values were plotted against its standard error in a funnel plot demonstrating asymmetry with a near-absence of studies reporting discrimination below 0.70 reference (Egger intercept 4.38, 95% CI: 3.61–5.16; $p < 0.001$; $n = 451$) (Figure 6). Funnel asymmetry has explanations besides publication bias – small-study effects, heterogeneity in task difficulty and the mathematical dependence between AUC and its variance. The finding is consistent with selective publication and selective outcome reporting of well-performing studies.

### 3.11. Clinical prediction task sub-stratification analysis

Thirty-five different clinical predictions tasks were identified. 9 tasks had >5 studies and were pooled. 16 tasks had studies between 2 and 5 and were described with median AUC and range, not pooled (Suppl. Table S8). Imaging-to-molecular tasks sit as the top performers (IDH mutation AUC 0.91, glioma grade AUC 0.90, MGMT methylation AUC 0.84) and clinical-outcome tasks at the bottom (readmission or reoperation AUC 0.82, seizure freedom after epilepsy surgery AUC 0.73). Glioma survival prediction task was the most numerous with 120 studies with 34 reporting AUC (0.83; 95% CI: 0.80–0.86) with only half reporting calibration. The spread across the tasks (AUC 0.73–0.91) is wider than the spread across subspecialities. Heterogeneity remained relevant for within-task analyses too ($\tau^2$:0.33 against 0.27 study-wide). Relatively homogenous pooled tasks were identified as seizure freedom after epilepsy surgery ($I^2$=22.2%, $\tau^2$=0.01) and shunt failure after hydrocephalus ($I^2$=77.2%, $\tau^2$=0.15).

## 4. Discussion

This is the largest corpus on the analysis of neurosurgical AI clinical prediction studies till date.[1–6] Sub-speciality level meta-analysis, sub-stratification prediction task analysis, risk-of-bias-stratified sensitivity analyses with prediction intervals and level of evidence synthesis are strengths of the study demonstrating heterogeneity.

The models reported uniformly high discrimination across all subspecialities (pooled AUC: 0.84–0.87), and this signal was robust to risk-of-bias stratification. Three findings temper the results. First, approximately nine in ten studies were at high overall risk of bias, largely because of the analysis domain. Second, heterogeneity was extreme ($I^2$~99%), so the pooled estimates summarise a highly dispersed literature rather than a transferable level of performance. Third, most models were internally validated and uncalibrated, so reported AUCs are best read as optimism-prone upper bounds.

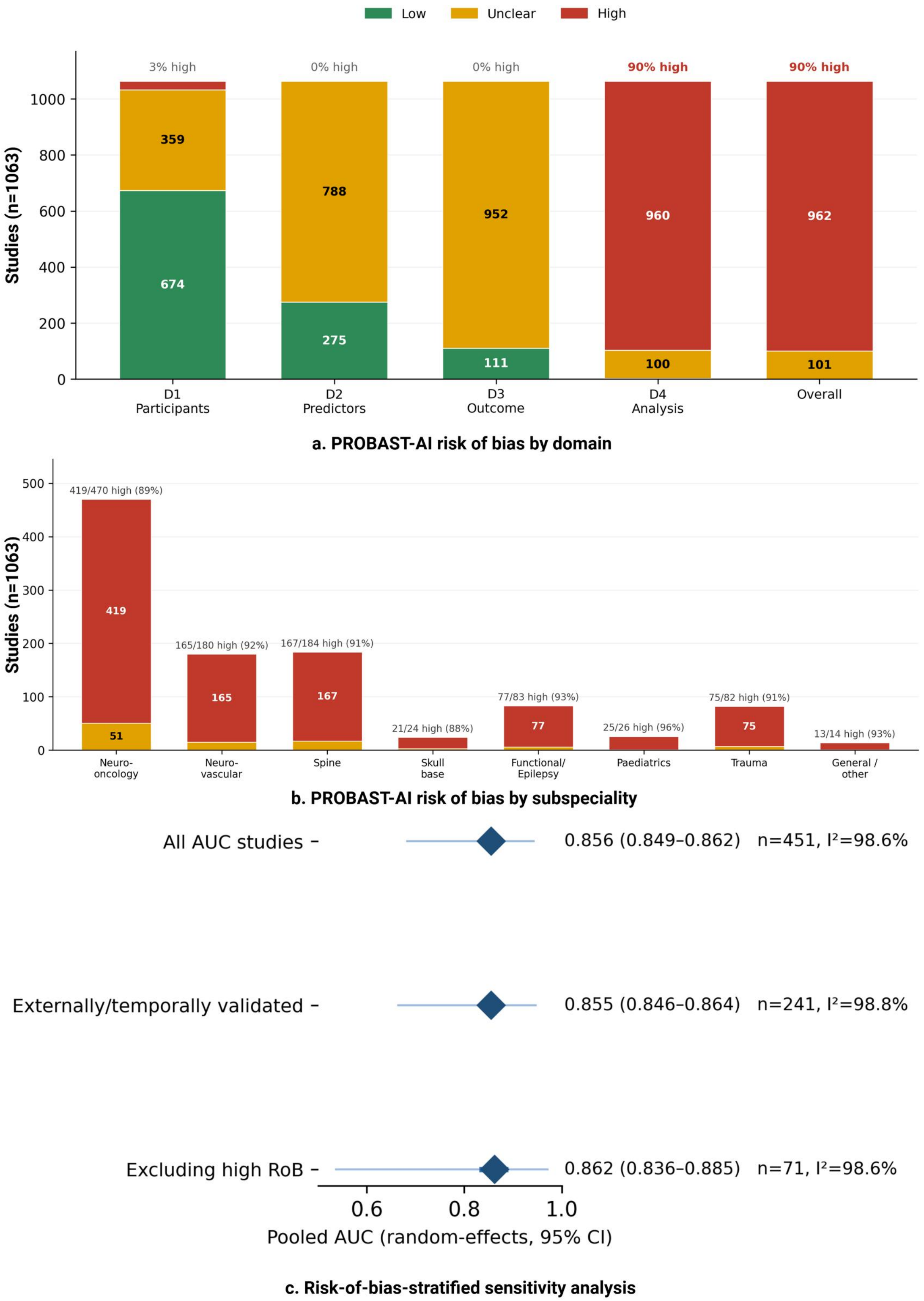

**Figure 5.** Risk-of-bias assessment using PROBAST checklist by domain (a), subspeciality (b)and risk-of-bias-stratified sensitivity analysis (c) for pooled AUC metric. There was overall high risk-of-bias overall dominated by analysis domain which was replicated by subspeciality risk-of-bias analysis with high risk-of-bias in each group ranging from 88% in skull base to 96% in paediatrics.
ROB: risk of bias; AUC: area under the curve; PROBAST: Prediction model Risk Of Bias ASsessment Tool

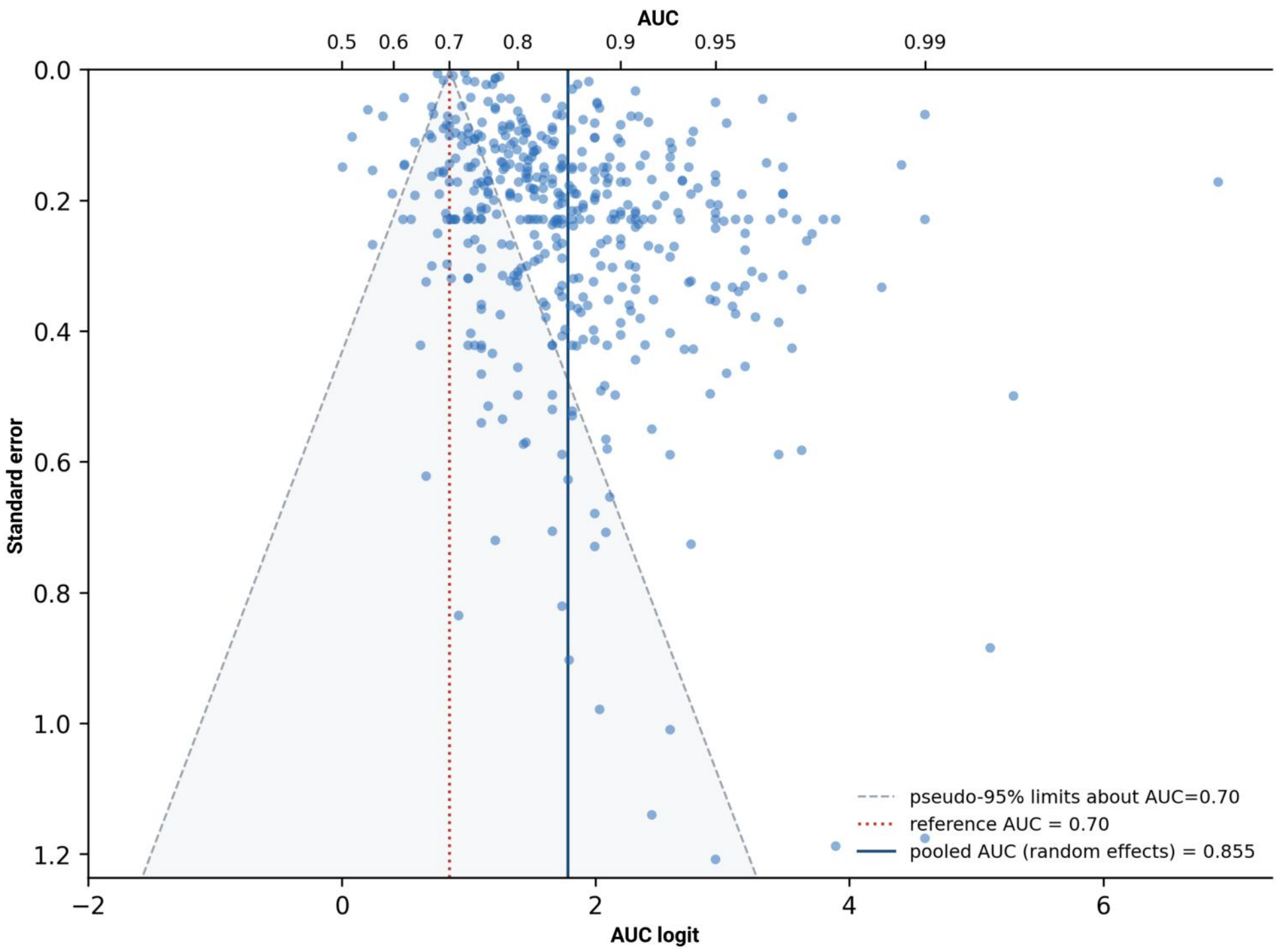

**Figure 6.** Funnel plot of AUC against its standard error for all AUC-reporting neurosurgical studies ($n = 451$).
Each point is one study's primary AUC, expressed on the logit scale (lower axis); the upper axis gives the corresponding AUC. The vertical axis is the standard error of the logit AUC, inverted. The red dotted line marks the reference AUC of 0.70 (conventional threshold for acceptable discrimination) about which the dashed pseudo-95% confidence limits and the shaded triangle are centred. The solid navy line is the random effects pooled estimate of AUC 0.86 (95% CI 0.849–0.862; $\tau^2 = 0.267$, $I^2 = 98.6\%$). Egger's regression test on the logit scale gave an intercept of 4.38 (95% CI 3.60–5.16), $p < 0.001$, indicating pronounced funnel asymmetry, with a conspicuous absence of studies reporting discrimination at or below the 0.70 reference.
AUC: area-under-the-curve; AUC logit: logit transformation of AUC scores for visualisation; standard error: measure of precision to detect asymmetry

Placing this study along the clinical neurosurgical care pathway sharpens where the field stands. Effort is heavily front-loaded at diagnosis, molecular characterisation and baseline prognostication especially the imaging-rich tasks where discrimination is highest. On the other hand, the stages closest to an operative decision are unevenly served: post-operative decision support and intraoperative guidance, remain sparse. These gaps and lack of treatment response modelling outside of neuro-oncology mark where clinically useful development is most needed and highest-value prediction (informing whether and how to operate or which therapy to give) is currently least mature.

The secondary metrics extracted independently of AUC and from a different subset of studies, reproduced the same ordering, with the diagnosis, grading and molecular stage with highest score (specificity 0.93) and prognostic/post-operative tasks with a lower score. Convergence across metrics derived from different reported quantities makes it unlikely that the clinical pathway gradient reflects selective AUC reporting alone and supports the interpretation that imaging-anchored classification is an easier prediction problem than forecasting a future clinical course.

Level of evidence assessment demonstrates the translation gap: only 5% of studies reach prospective cohort or trial data with external or temporal validation, while 58% rest on internal validation alone or on apparent performance with no validation. Level of evidence and risk of bias were aligned: every internal or no validation study was rated high risk of bias identifying the deficits in these studies. The subspeciality with the strongest discrimination (neuro-oncology) had the lowest proportion of prospective evidence, whereas neurovascular neurosurgery, with more prospective and externally validated work, reported lower pooled AUCs. This inverse relationship shows that apparent performance is inflated by optimism and is a caution against interpreting higher pooled AUC as evidence of greater clinical readiness.

These patterns mirror the broader literature on clinical ML prediction, where the analysis domain is universally the weak point. The recurring, addressable deficiencies were absent calibration assessment (discrimination-only reporting), internal-only validation and inconsistent handling and reporting of missing data. Small number of studies that are externally validated and non-high-risk subsets produced similar high pooled AUCs demonstrating that performance is not being increased solely by the weakest work, but the persistence of heterogeneity shows that aggregate discrimination is not a meaningful guide to the performance of individual models in a new unseen population.

The practical implication is that the neurosurgical field[1–6] should shift emphasis from reporting increasing in-sample AUCs and secondary metrics towards demonstrating external and temporal validation, routine calibration, adequate events-per-variable, transparent missing-data handling and complete reporting of these studies using standardised reporting guidelines (TRIPOD, TRIPOD-AI, PROBAST, PROBAST-AI)[13–15] that are prerequisites for trustworthy clinical deployment.

### 4.1. The translation gap

Across all studies spanning every neurosurgical subspeciality and a decade of publication, 63.9% studies stopped at model development and a further 31.2% at external validation. Only 52 studies (4.9%) produced an artefact a clinician could use or evaluated the model in prospective clinical practice, and none reported an impact or implementation study. The field is generating discrimination statistics at scale but not enough clinical evidence for neurosurgical application. The bottleneck is no longer algorithmic; it is calibration at local population, a usable interface, prospective silent period testing and a design capable of detecting whether neurosurgical decisions and outcomes change.

## 5. Limitations

### 5.1. Studies assessed

Limitations encompassed three main areas.

First, pooling of single AUC scores per study across heterogeneous outcomes, predictors and population is coarse, made visible by the extreme heterogeneity $I^2$ and $\tau^2$ values. AUC metric alone can be misrepresentative and thus performance assessment should be done based on multiple complementary metrics and task-based sub-stratification analyses. Second, secondary metrics were pooled only from studies that reported a single value for each metric leading to these studies being small in number and may not be representative and paired data were limited to permit a bivariate modelling. Third, in minority of studies that did not have CI and sample size reported, standard errors were imputed, so pooled CI are approximate.

### 5.2. Review process

Title and abstract level exclusion of studies is likely to provide a conservative method in including ML prediction studies for neurosurgery. Publication bias may have affected the range of metrics and studies included in the systematic review especially by exclusion of preprints and materials that have not undergone peer review partially answered by small study effects and publication bias analyses. This is relevant to these studies where the rapid pace of development often outstrips the slower process of peer review, leading to some researchers avoiding the submission to peer-reviewed journals.

## 6. Conclusion

AI based clinical prediction tools in neurosurgery report consistently high discrimination across subspecialities and clinical predictions tasks, but the literature is dominated by internally validated and poorly calibrated models with extreme heterogeneity with majority at high risk of bias. External and temporal validation, calibrating reporting and adherence to reporting guidelines are immediate priorities to translate promising discrimination into trustworthy clinical tools in neurosurgery.

## Author contributions

PG conceived the study, performed search, screening, extraction, analysis and drafting. Co-authors AC, JS, KA contributed to second review, interpretation and critical revision. All authors reviewed and approved the manuscript.

## Ethical statement

All studies were retrieved from published literature. No explicit ethical approval was obtained.

## Disclosure statement

No potential conflict of interest was reported by the author(s).

## Software and AI-use declaration

RStudio, Python/PyCharm and Biorender.com were used to perform statistical analyses and create graphs respectively. Authors declare no use of AI during drafting of the manuscript or any supplementary materials.

## ORCID

Prajwal Ghimire http://orcid.org/0000-0003-2884-956X

## Data availability

Full study level datasets, metrics calculations and PROBAST-AI domain judgements are available as supplemental files.